\documentclass[aps,prl,showpacs,twocolumn,groupedaddress]{revtex4}  
\usepackage{graphicx}
\usepackage{dcolumn}
\usepackage{bm}
\usepackage{epsfig}

\begin{document}

%
%
\newcommand{\comm}[1]{\mbox{\mbox{\textup{#1}}}}
\newcommand{\subs}[1]{\mbox{\scriptstyle \mathit{#1}}}
\newcommand{\subss}[1]{\mbox{\scriptscriptstyle \mathit{#1}}}
\newcommand{\Frac}[2]{\mbox{\frac{\displaystyle{#1}}{\displaystyle{#2}}}}
\newcommand{\LS}[1]{\mbox{_{\scriptstyle \mathit{#1}}}}
\newcommand{\US}[1]{\mbox{^{\scriptstyle \mathit{#1}}}}
\def\gsim{\mathrel{\rlap{\raise.4ex\hbox{$>$}} {\lower.6ex\hbox{$\sim$}}}}
\def\lsim{\mathrel{\rlap{\raise.4ex\hbox{$<$}} {\lower.6ex\hbox{$\sim$}}}}
\renewcommand{\arraystretch}{1.3}
\newcommand{\Edep}{\mbox{E_{\mathit{dep}}}}
\newcommand{\Ebeam}{\mbox{E_{\mathit{beam}}}}
\newcommand{\Exrc}{\mbox{E_{\mathit{rec}}^{\mathit{exp}}}}
\newcommand{\Emrc}{\mbox{E_{\mathit{rec}}^{\mathit{sim}}}}
\newcommand{\Evis}{\mbox{E_{\mathit{vis}}}}
\newcommand{\Edepi}{\mbox{E_{\mathit{dep,i}}}}
\newcommand{\Evisi}{\mbox{E_{\mathit{vis,i}}}}
\newcommand{\Exrci}{\mbox{E_{\mathit{rec,i}}^{\mathit{exp}}}}
\newcommand{\Emrci}{\mbox{E_{\mathit{rec,i}}^{\mathit{sim}}}}
\newcommand{\Etmis}{\mbox{E_{\mathit{t,miss}}}}
%
%
\newcommand{\lt}{\mbox{$<$}}
\newcommand{\gt}{\mbox{$>$}}
\newcommand{\lte}{\mbox{$<=$}}
\newcommand{\gte}{\mbox{$>=$}}

\newcommand{\xa}{\mbox{$x_{a}$}}
\newcommand{\xb}{\mbox{$x_{b}$}}
\newcommand{\xp}{\mbox{$x_{p}$}}
\newcommand{\xpb}{\mbox{$x_{\bar{p}}$}}

\newcommand{\alphas}{\mbox{$\alpha_{s}$}}
\newcommand{\partt}{\mbox{$\partial^{t}$}}
\newcommand{\partd}{\mbox{$\partial_{t}$}}
\newcommand{\parmut}{\mbox{$\partial^{\mu}$}}
\newcommand{\parmud}{\mbox{$\partial_{\mu}$}}
\newcommand{\parnut}{\mbox{$\partial^{\nu}$}}
\newcommand{\parnud}{\mbox{$\partial_{\nu}$}}
\newcommand{\Amut}{\mbox{$A^{\mu}$}}
\newcommand{\Amud}{\mbox{$A_{\mu}$}}
\newcommand{\Anut}{\mbox{$A^{\nu}$}}
\newcommand{\Anud}{\mbox{$A_{\nu}$}}
\newcommand{\Fmunut}{\mbox{$F^{\mu\nu}$}}
\newcommand{\Fmunud}{\mbox{$F_{\mu\nu}$}}
\newcommand{\Gmut}{\mbox{$\gamma^{\mu}$}}
\newcommand{\Gmud}{\mbox{$\gamma_{\mu}$}}
\newcommand{\Gnut}{\mbox{$\gamma^{\nu}$}}
\newcommand{\Gnud}{\mbox{$\gamma_{\nu}$}}
\newcommand{\pvecsq}{\mbox{$\vec{p}^{\:2}$}}
\newcommand{\bigsum}{\mbox{$\displaystyle{\sum}$}}
%
\newcommand{\Lzero}{Level \O}
\newcommand{\Lone}{Level $1$}
\newcommand{\Ltwo}{Level $2$}
\newcommand{\Lhalf}{Level $1.5$}
%
\newcommand{\bzero}{\mbox{\comm{B\O}}}
\newcommand{\dzero}{\mbox{D\O}}
\newcommand{\dzerosm}{\mbox{\comm{$\scriptsize{D\O}$}}}
\newcommand{\runb}{Run~$1$B}
\newcommand{\runa}{Run~$1$A}
\newcommand{\runone}{Run~$1$}
\newcommand{\runtwo}{Run~$2$}
\newcommand{\dzpjet}{\textsc{D{\O}Pjet}}
\newcommand{\y}{\mbox{$y$}}
\newcommand{\z}{\mbox{$z$}}
\newcommand{\px}{\mbox{$p_{x}$}}
\newcommand{\py}{\mbox{$p_{y}$}}
\newcommand{\pz}{\mbox{$p_{z}$}}
\newcommand{\ex}{\mbox{$E_{x}$}}
\newcommand{\ey}{\mbox{$E_{y}$}}
\newcommand{\ez}{\mbox{$E_{z}$}}
\newcommand{\et}{\mbox{$E_{T}$}}
\newcommand{\etprime}{\mbox{$E_{T}^{\prime}$}}
\newcommand{\etone}{\mbox{$E_{T}^{\mathrm{1}}$}}
\newcommand{\ettwo}{\mbox{$E_{T}^{\mathrm{2}}$}}
\newcommand{\etlj}{\mbox{$E_{T}^{\subs{lj}}$}}
\newcommand{\etmax}{\mbox{$E_{T}^{max}$}}
\newcommand{\etcand}{\mbox{$E_{T}^{\subs{cand}}$}}
\newcommand{\etup}{\mbox{$E_{T}^{\subs{up}}$}}
\newcommand{\etdown}{\mbox{$E_{T}^{\subs{down}}$}}
\newcommand{\jet}{\mbox{$E_{T}^{\subs{jet}}$}}
\newcommand{\cet}{\mbox{$E_{T}^{\subs{cell}}$}}
\newcommand{\jetvec}{\mbox{$\vec{E}_{T}^{\subs{jet}}$}}
\newcommand{\cetvec}{\mbox{$\vec{E}_{T}^{\subs{cell}}$}}
\newcommand{\jevec}{\mbox{$\vec{E}^{\subs{jet}}$}}
\newcommand{\cevec}{\mbox{$\vec{E}^{\subs{cell}}$}}
\newcommand{\etfr}{\mbox{$f_{E_{T}}$}}
\newcommand{\aveet}{\mbox{$\langle\et\rangle$}}
\newcommand{\nj}{\mbox{$n_{j}$}}
\newcommand{\ptrel}{\mbox{$p_{T}^{rel}$}}
\newcommand{\etad}{\mbox{$\eta_{d}$}}           
\newcommand{\peta}{\mbox{$\eta$}}               
\newcommand{\aeta}{\mbox{$|\eta|$}}             
\newcommand{\ifb}{fb$^{-1}$}
\newcommand{\ipb}{pb$^{-1}$}
\newcommand{\inb}{nb$^{-1}$}
\newcommand{\met}{\mbox{${\hbox{$E$\kern-0.63em\lower-.18ex\hbox{/}}}_{T}$}}
\newcommand{\metvec}{\mbox{${\hbox{$\vec{E}$\kern-0.63em\lower-.18ex\hbox{/}}}_{T}\,$}}
\newcommand{\metx}{\mbox{${\hbox{$E$\kern-0.63em\lower-.18ex\hbox{/}}}_{x}\,$}}
\newcommand{\mety}{\mbox{${\hbox{$E$\kern-0.63em\lower-.18ex\hbox{/}}}_{y}\,$}}
\newcommand{\het}{\mbox{$\vec{\mathcal{H}}_{T}$}}
\newcommand{\hetsc}{\mbox{$\mathcal{H}_{T}$}}
\newcommand{\zvrt}{\mbox{$Z$}}
\newcommand{\zcut}{\mbox{$|\zvrt| < 50$}}
\newcommand{\mtwo}{\mbox{$\mathcal{M}_{2}$}}
\newcommand{\mthree}{\mbox{$\mathcal{M}_{3}$}}
\newcommand{\mfour}{\mbox{$\mathcal{M}_{4}$}}
\newcommand{\msix}{\mbox{$\mathcal{M}_{6}$}}
\newcommand{\mn}{\mbox{$\mathcal{M}_{n}$}}
\newcommand{\R}{\mbox{$R_{\subss{MTE}}$}}
\newcommand{\invR}{\mbox{$1/R_{\subss{MTE}}$}}
\newcommand{\eemf}{\mbox{$\varepsilon_{\subss{EMF}}$}}
\newcommand{\echf}{\mbox{$\varepsilon_{\subss{CHF}}$}}
\newcommand{\ehcf}{\mbox{$\varepsilon_{\subss{HCF}}$}}
\newcommand{\eglob}{$\mbox{\varepsilon_{\subs{glob}}$}}
\newcommand{\emte}{\mbox{$\varepsilon_{\subss{MTE}}$}}
\newcommand{\ezvrt}{\mbox{$\varepsilon_{\subss{Z}}$}}
\newcommand{\etot}{\mbox{$\varepsilon_{\subs{tot}}$}}
\newcommand{\Bprime}{\mbox{$\comm{B}^{\prime}$}}
\newcommand{\Nsurv}{\mbox{$N_{\subs{surv}}$}}
\newcommand{\Nfail}{\mbox{$N_{\subs{fail}}$}}
\newcommand{\Ntot}{\mbox{$N_{\subs{tot}}$}}
\newcommand{\p}[1]{\mbox{$p_{#1}$}}
\newcommand{\ep}[1]{\mbox{$\Delta p_{#1}$}}
\newcommand{\delr}{\mbox{$\Delta R$}}
\newcommand{\deleta}{\mbox{$\Delta\eta$}}
\newcommand{\cafone}{{\sc Cafix 5.1}}
\newcommand{\caftwo}{{\sc Cafix 5.2}}
\newcommand{\delphi}{\mbox{$\Delta\varphi$}}
\newcommand{\rphi}{\mbox{$r-\varphi$}}
\newcommand{\etaphi}{\mbox{$\eta-\varphi$}}
\newcommand{\etatphi}{\mbox{$\eta\times\varphi$}}
\newcommand{\Rjet}{\mbox{$R_{jet}$}}
\newcommand{\jphi}{\mbox{$\varphi_{\subs{jet}}$}}
\newcommand{\gphi}{\mbox{$\varphi_{\subs{\gamma}}$}}
\newcommand{\ceta}{\mbox{$\eta^{\subs{cell}}$}}
\newcommand{\cphi}{\mbox{$\phi^{\subs{cell}}$}}
\newcommand{\inlum}{\mbox{$\mathcal{L}$}}
\newcommand{\gm}{\mbox{$\gamma$}}
\newcommand{\Rjj}{\mbox{$\mathbf{R}_{\subs{jj}}$}}
\newcommand{\Rgj}{\mbox{$\mathbf{R}_{\subs{\gamma j}}$}}
\newcommand{\Rmathcal}{\mbox{$\mathcal{R}$}}
\newcommand{\etv}{\mbox{$\vec{E}_{T}$}}
\newcommand{\nvec}{\mbox{$\hat{\vec{n}}$}}
\newcommand{\eprime}{\mbox{$E^{\prime}$}}
\newcommand{\aveprime}{\mbox{$\bar{E}^{\prime}$}}
\newcommand{\geta}{\mbox{$\eta_{\gm}$}}
\newcommand{\jeta}{\mbox{$\eta_{\subs{jet}}$}}
\newcommand{\cjeta}{\mbox{$\eta_{\subs{jet}}^{\subss{CEN}}$}}
\newcommand{\fjeta}{\mbox{$\eta_{\subs{jet}}^{\subss{FOR}}$}}
\newcommand{\cjphi}{\mbox{$\varphi_{\subs{jet}}^{\subss{CEN}}$}}
\newcommand{\fjphi}{\mbox{$\varphi_{\subs{jet}}^{\subss{FOR}}$}}
\newcommand{\etcut}{\mbox{$E_{T}^{\subs{cut}}$}}
\newcommand{\etg}{\mbox{$E_{T}^{\gm}$}}
\newcommand{\cenet}{\mbox{$E_{T}^{\subss{CEN}}$}}
\newcommand{\foret}{\mbox{$E_{T}^{\subss{FOR}}$}}
\newcommand{\cenen}{\mbox{$E^{\subss{CEN}}$}}
\newcommand{\foren}{\mbox{$E^{\subss{FOR}}$}}
\newcommand{\ejtptc}{\mbox{$E^{\subs{ptcl}}_{\subs{jet}}$}}
\newcommand{\ejtmes}{\mbox{$E^{\subs{meas}}_{\subs{jet}}$}}
\newcommand{\AIDA}{{\sc AIDA}}
\newcommand{\RECO}{{\sc Reco}}
\newcommand{\PYTHIA}{{\sc Pythia}}
\newcommand{\HERWIG}{{\sc Herwig}}
\newcommand{\JETRAD}{{\sc Jetrad}}
\newcommand{\CTone}{\mbox{$|\eta|<0.4}$}
\newcommand{\CTtwo}{\mbox{$0.4\leq|\eta|<0.8$}}
\newcommand{\ICone}{\mbox{$0.8\leq|\eta|<1.2$}}
\newcommand{\ICtwo}{\mbox{$1.2\leq|\eta|<1.6$}}
\newcommand{\FWone}{\mbox{$1.6\leq|\eta|<2.0$}}
\newcommand{\FWtwo}{\mbox{$2.0\leq|\eta|<2.5$}}
\newcommand{\FWthr}{\mbox{$2.5\leq|\eta|<3.0$}}
\newcommand{\LCTone}{\mbox{$|\eta|<0.5$}}
\newcommand{\LCTtwo}{\mbox{$0.5\leq|\eta|<1.0$}}
\newcommand{\LICone}{\mbox{$1.0\leq|\eta|<1.5$}}
\newcommand{\LICtwo}{\mbox{$1.5\leq|\eta|<2.0$}}
\newcommand{\LFWone}{\mbox{$2.0\leq|\eta|<3.0$}}
\newcommand{\CSone}{\mbox{$|\eta|<0.5$}}
\newcommand{\CStwo}{\mbox{$0.5\leq|\eta|<1.0$}}
\newcommand{\CSthr}{\mbox{$1.0\leq|\eta|<1.5$}}
\newcommand{\CSfou}{\mbox{$1.5\leq|\eta|<2.0$}}
\newcommand{\CSfiv}{\mbox{$2.0\leq|\eta|<3.0$}}
\newcommand{\sigA}{\mbox{$\sigma_{\subss{\!A}}$}}
\newcommand{\sigASS}{\mbox{$\sigma_{\subss{A}}^{\subss{SS}}$}}
\newcommand{\sigAOS}{\mbox{$\sigma_{\subss{A}}^{\subss{OS}}$}}
\newcommand{\sigZ}{\mbox{$\sigma_{\subss{Z}}$}}
\newcommand{\sige}{\mbox{$\sigma_{\subss{E}}$}}
\newcommand{\siget}{\mbox{$\sigma_{\subss{\et}}$}}
\newcommand{\sigetone}{\mbox{$\sigma_{\subs{\etone}}$}}
\newcommand{\sigettwo}{\mbox{$\sigma_{\subs{\ettwo}}$}}
\newcommand{\rcal}{\mbox{$R_{\subs{cal}}$}}
\newcommand{\zcal}{\mbox{$Z_{\subs{cal}}$}}
\newcommand{\Runf}{\mbox{$R_{\subs{unf}}$}}
\newcommand{\Rsep}{\mbox{$\mathcal{R}_{sep}$}}
\newcommand{\etal}{{\it et al.}}
\newcommand{\ppbar}{\mbox{$p\overline{p}$}}
\newcommand{\pp}{\mbox{$pp$}}
\newcommand{\qqbar}{\mbox{$q\overline{q}$}}
\newcommand{\ccbar}{\mbox{$c\overline{c}$}}
\newcommand{\bbbar}{\mbox{$b\overline{b}$}}
\newcommand{\ttbar}{\mbox{$t\overline{t}$}}

\newcommand{\bbj}{\mbox{$b\overline{b}j$}}
\newcommand{\bbjj}{\mbox{$b\overline{b}jj$}}
\newcommand{\ccjj}{\mbox{$c\overline{c}jj$}}

\newcommand{\bb}{\mbox{$b\overline{b}j(j)$}}
\newcommand{\cc}{\mbox{$c\overline{c}j(j)$}}

\newcommand{\hboson}{\mbox{$\mathit{h}$}}
\newcommand{\Hboson}{\mbox{$\mathit{H}$}}
\newcommand{\Aboson}{\mbox{$\mathit{A}$}}
\newcommand{\zboson}{\mbox{$\mathit{Z}$}}
\newcommand{\zb}{\mbox{$\mathit{Zb}$}}
\newcommand{\bh}{\mbox{$\mathit{bh}$}}
\newcommand{\btag}{\mbox{$\mathit{b}$}}

\newcommand{\hsm}{\mbox{$h_{SM}$}}
\newcommand{\hmssm}{\mbox{$h_{MSSM}$}}

\newcommand{\prot}{\mbox{$p$}}
\newcommand{\pbar}{\mbox{$\overline{p}$}}
\newcommand{\pt}{\mbox{$p_{T}$}}
\newcommand{\xnot}{\mbox{$X_{0}$}}
\newcommand{\Znot}{\mbox{$Z^{0}$}}
\newcommand{\Wpm}{\mbox{$W^{\pm}$}}
\newcommand{\Wplus}{\mbox{$W^{+}$}}
\newcommand{\Wminus}{\mbox{$W^{-}$}}
\newcommand{\lamb}{\mbox{$\lambda$}}
\newcommand{\nhatbf}{\mbox{$\hat{\mathbf{n}}$}}
\newcommand{\pbf}{\mbox{$\mathbf{p}$}}
\newcommand{\xbf}{\mbox{$\mathbf{x}$}}
\newcommand{\jbf}{\mbox{$\mathbf{j}$}}
\newcommand{\Ebf}{\mbox{$\mathbf{E}$}}
\newcommand{\Bbf}{\mbox{$\mathbf{B}$}}
\newcommand{\Abf}{\mbox{$\mathbf{A}$}}
\newcommand{\Rbf}{\mbox{$\mathbf{R}$}}
\newcommand{\nablabf}{\mbox{$\mathbf{\nabla}$}}
\newcommand{\rarrow}{\mbox{$\rightarrow$}}
\newcommand{\slashp}{\mbox{$\not \! p \,$}}
\newcommand{\slashk}{\mbox{$\not \! k$}}
\newcommand{\slasha}{\mbox{$\not \! a$}}
\newcommand{\slashA}{\mbox{$\! \not \! \! A$}}
\newcommand{\slashpar}{\mbox{$\! \not \! \partial$}}
\newcommand{\intdouble}{\mbox{$\int\!\!\int$}}
\newcommand{\MRSTGU}{MRSTg$\uparrow$}
\newcommand{\MRSTGD}{MRSTg$\downarrow$}
%
\newcommand{\Due}{\mbox{$D_{\mathrm{ue}}$}}
\newcommand{\Dth}{\mbox{$D_{\Theta}$}}
\newcommand{\Dof}{\mbox{$D_{\mathrm{O}}$}}
\newcommand{\zbl}{\texttt{ZERO BIAS}}
\newcommand{\mbl}{\texttt{MIN BIAS}}
\newcommand{\mbll}{\texttt{MINIMUM BIAS}}
\newcommand{\nue}{\mbox{$\nu_{e}$}}
\newcommand{\num}{\mbox{$\nu_{\mu}$}}
\newcommand{\nut}{\mbox{$\nu_{\tau}$}}
\newcommand{\mycs}{\mbox{$d^{\,2}\sigma/(d\et d\eta)$}}
\newcommand{\mycsav}{\mbox{$\langle \mycs \rangle$}}
\newcommand{\tdcs}{\mbox{$d^{\,3}\sigma/d\et d\eta_{1} d\eta_{2}$}}
\newcommand{\tdcsav}{\mbox{$\langle d^{\,3}\sigma/d\et d\eta_{1} d\eta_{2} \rangle$}}
\newcommand{\tanb}{$\tan\beta$}
\newcommand{\cotb}{$\cot\beta$}

\newcommand{\rstev}{\mbox{$\rs = \T{1.8}$}}
\newcommand{\rssps}{\mbox{$\rs = \T{0.63}$}}
\newcommand{\XX}{\mbox{$\, \times \,$}}
\newcommand{\AP}{\mbox{${\rm \bar{p}}$}}
\newcommand{\SU}{\mbox{$<\! |S|^2 \!>$}}
\newcommand{\ET}{\mbox{$E_{T}$}}
\newcommand{\HT}{\mbox{$S_{{\rm {\sl T}}}$} }
\newcommand{\PT}{\mbox{$p_{t}$}}
\newcommand{\DP}{\mbox{$\Delta\phi$}}
\newcommand{\DR}{\mbox{$\Delta R$}}
\newcommand{\DE}{\mbox{$\Delta\eta$}}
\newcommand{\DEP}{\mbox{$\Delta\eta_{c}$}}
\newcommand{\PH}{\mbox{$\phi$}}
\newcommand{\EA}{\mbox{$\eta$} }
\newcommand{\EAJ}{\mbox{\EA(jet)}}
\newcommand{\AEA}{\mbox{$|\eta|$}}
\newcommand{\Ge}[1]{\mbox{#1 GeV}}
\newcommand{\T}[1]{\mbox{#1 TeV}}
\newcommand{\x}{\cdot}
\newcommand{\ra}{\rightarrow}
\def\D0{D\O}
\def\ETmiss{{\rm {\mbox{$E\kern-0.57em\raise0.19ex\hbox{/}_{T}$}}}}
\newcommand{\mb}{\mbox{mb}}
\newcommand{\nb}{\mbox{nb}}
\newcommand{\rs}{\mbox{$\sqrt{\rm {\sl s}}$}}
\newcommand{\fdel}{\mbox{$f(\DEP)$}}
\newcommand{\fdele}{\mbox{$f(\DEP)^{exp}$}}
\newcommand{\fgap}{\mbox{$f(\DEP\! \geq \!3)$}}
\newcommand{\fgape}{\mbox{$f(\DEP\! \geq \!3)^{exp}$}}
\newcommand{\fpyt}{\mbox{$f(\DEP\!>\!2)$}}
\newcommand{\delth}{\mbox{$\DEP\! \geq \!3$}}
\newcommand{\uplim}{\mbox{$1.1\!\times\!10^{-2}$}}
\def\simge
{\mathrel{\rlap{\raise 0.53ex \hbox{$>$}}{\lower 0.53ex \hbox{$\sim$}}}}
\def\simle
{\mathrel{\rlap{\raise 0.53ex \hbox{$<$}}{\lower 0.53ex \hbox{$\sim$}}}}
\newcommand{\pbarp}{\mbox{$p\bar{p}$}}
\def\ETmiss{\mbox{${\hbox{$E$\kern-0.5em\lower-.1ex\hbox{/}\kern+0.15em}}_{\rm T}$}}
\def\Et{\mbox{$E_{T}$}}
\newcommand{\modeta}{\mid \!\! \eta \!\! \mid}
\def\gevcc{GeV/c$^2$}                   
\def\gevc{GeV/c}                        
\def\gev{GeV}                           
\newcommand{\als}{\mbox{${\alpha_{{\rm s}}}$}}
\def\1960{$\sqrt{s}=1960$ GeV}
\def\etI{E_{T_1}}
\def\etII{E_{T_2}}
\def\itaI{\eta_1}
\def\itaII{\eta_2}
\def\deta{\Delta\eta}
\def\etab{\bar{\eta}}
\def\xq{($x_1$,$x_2$,$Q^2$)}
\def\xx{($x_1$,$x_2$)}
\def\rap{pseudorapidity}
\def\as{\alpha_s}
\def\ap{\alpha_{\rm BFKL}}
\def\apb{\alpha_{{\rm BFKL}_{bin}}}
\def\cm{c.m.}




\hspace{5.2in}\mbox{FERMILAB-PUB-09-299-E} 

\title{Search for resonant pair production of neutral long-lived particles
decaying to {\boldmath \bbbar} in {\boldmath \ppbar} collisions at
{\boldmath $\sqrt{s}=1.96$}~TeV}

%
\author{V.M.~Abazov$^{37}$}
\author{B.~Abbott$^{75}$}
\author{M.~Abolins$^{65}$}
\author{B.S.~Acharya$^{30}$}
\author{M.~Adams$^{51}$}
\author{T.~Adams$^{49}$}
\author{E.~Aguilo$^{6}$}
\author{M.~Ahsan$^{59}$}
\author{G.D.~Alexeev$^{37}$}
\author{G.~Alkhazov$^{41}$}
% ...

\affiliation{\vspace{0.1 in}(The D\O\ Collaboration)\vspace{0.1 in}}
% ...
\affiliation{$^{6}$University of Alberta, Edmonton, Alberta, Canada;
                Simon Fraser University, Burnaby, British Columbia, Canada;
                York University, Toronto, Ontario, Canada and
                McGill University, Montreal, Quebec, Canada}
% ...
\affiliation{$^{30}$Tata Institute of Fundamental Research, Mumbai, India}
% ...
\affiliation{$^{37}$Joint Institute for Nuclear Research, Dubna, Russia}
% ...
\affiliation{$^{41}$Petersburg Nuclear Physics Institute,
                St. Petersburg, Russia}
% ...
\affiliation{$^{49}$Florida State University, Tallahassee, Florida 32306, USA}
% ...
\affiliation{$^{51}$University of Illinois at Chicago,
                Chicago, Illinois 60607, USA}
% ...
\affiliation{$^{59}$Kansas State University, Manhattan, Kansas 66506, USA}
% ...
\affiliation{$^{65}$Michigan State University,
                East Lansing, Michigan 48824, USA}
% ...
\affiliation{$^{75}$University of Oklahoma, Norman, Oklahoma 73019, USA}
% ...

\date{June 9, 2009}

\begin{abstract}
We report on a first search for resonant pair production of
neutral long-lived particles (NLLP) which each decay to a
\bbbar\ pair, using 3.6 \ifb\ of data recorded with the D0 detector
at the Fermilab Tevatron collider. We search for pairs of displaced
vertices in the tracking detector at radii in the range 1.6--20~cm
from the beam axis. No significant excess is observed above
background, and upper limits are set on the production rate in a
hidden-valley benchmark model for a range of Higgs boson masses and
NLLP masses and lifetimes.
\end{abstract}


\maketitle

A class of hidden-valley (HV) models \cite{strass2006_1} predicts a
new, confining gauge group that is weakly coupled to the standard
model (SM), leading to the production of HV particles (v-particles).
The details of v-particle decay depend on the specific model, but
the HV quarks always hadronize due to confinement producing
``v-hadrons'' that can be long-lived. One particular model used as a
benchmark for this search is the SM Higgs boson ($H$) mixing with a
HV Higgs boson that gives mass to v-particles. The SM Higgs boson
could then decay directly to v-hadrons through this mixing with a
substantial branching fraction~\cite{strass2006_2}. These v-hadrons
may couple preferentially to heavy SM particles, such as $b$ quarks,
due to helicity suppression. The result is a striking experimental
signature of highly displaced secondary vertices (SV) with a large
number of attached tracks from the $b$ quark decays. Direct searches
at the CERN LEP collider have excluded a Higgs boson decaying to
$b\bar{b}$ or $\tau\bar{\tau}$ with $M_{H}<114.4$ GeV at the 95\%
C.L.~\cite{lep2003}. But if the Higgs boson dominantly decays to
long-lived v-particles which then decay inside the detector to
\bbbar, only the most general LEP limit is relevant, $M_{H}>81$ GeV,
for any Higgs boson radiating off a $Z$ boson
\cite{spencer_higgs_decays}. Cosmological constraints require that
one of the light v-hadrons have a lifetime $\ll1$~second to be
consistent with models of big bang
nucleosynthesis~\cite{strass2006_1}.

In this Letter, we present the first search for pair-produced
neutral long-lived particles (NLLP), each decaying to a $b$ quark
pair, using the D0 detector \cite{d0det} at the Fermilab Tevatron
\ppbar\ collider. The $b$ quarks are required in order to provide a
high transverse momentum (\pt) muon for triggering with high
efficiency. The data were collected from April 2002 to August 2008
and correspond to an integrated luminosity of 3.6~\ifb\ at
$\sqrt{s}=1.96$~TeV. The D0 central tracking detector comprises a
silicon microstrip tracker (SMT) and a central fiber tracker (CFT),
both located within a 2~T superconducting solenoidal magnet. The
SMT, extending from a radius of $\approx$2~cm to $\approx$10~cm, has
a six-barrel longitudinal structure, each with a set of four layers
arranged axially around the beam pipe, and interspersed with 16
radial disks. The CFT, extending from a radius of $\approx$20~cm to
$\approx$50~cm, has eight thin coaxial barrels, each supporting two
doublets of overlapping scintillating fibers. Secondary vertices are
reconstructed by combining charged particle tracks found in the
tracking detector, which effectively limits the analysis to NLLP
decays occurring within a maximum radius of 20~cm, well within the
tracker volume. We also exclude vertex radii less than 1.6~cm since
the background from heavy flavor production is large in that region.
Known sources of SVs other than heavy-flavor include decays
in-flight of light particles, inelastic interactions of particles
with nuclei of detector material, and photon conversions. Vertices
may also be mimicked by pattern recognition errors.

\begin{figure}
\begin{centering}
\includegraphics[width=3.25in]{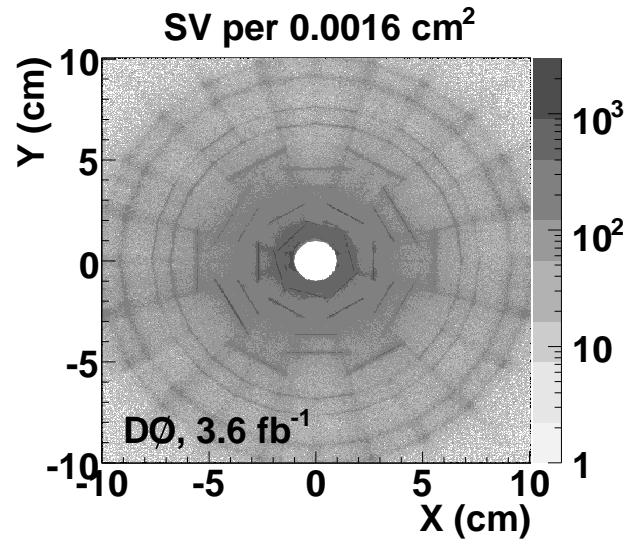}
\caption{\label{fig:xy-material-map}Material map in the plane
transverse to the beamline, generated using SVs with three attached
tracks, in events passing the initial selection. The structure of
the silicon detector and supports are clearly seen.}
\end{centering}
\end{figure}

{\sc pythia}~\cite{pythia} is used to simulate signal and background
events, which are then passed through a full {\sc
geant}3-based~\cite{geant} D0 detector simulation and the same
reconstruction as for collider data. For signal, the SM
$gg$\rarrow$H$ process is generated, the Higgs boson is forced to
decay to a pair of long-lived $A$ bosons (a heavy, neutral scalar,
representing a v-hadron), and each $A$ boson is forced to decay to a
pair of $b$ quarks. The Higgs boson mass ($M_H$) is varied from 90
to 200~GeV, the v-hadron mass ($m_{HV}$) from 15 to 40~GeV and the
average v-hadron proper decay length ($L_d = c\tau$) from 2.5~cm to
10~cm. For background, inclusive \ppbar\ multijet events are
generated. Approximately one hundred thousand Monte Carlo (MC)
events for each signal sample and ten million events of multijet
background are generated and are overlaid with data to simulate
detector noise and pile-up effects from additional \ppbar\
interactions.

At least two jets with a cone radius of 0.5~\cite{RunIIcone} are
required, each with $p_{T}>10$ GeV. And at least one muon is
required with $p_{T}>4$ GeV, matched within $\Delta {\cal R}<0.7$ to
one of the jets, where $\Delta {\cal R} = \sqrt{(\Delta\phi)^{2} +
(\Delta\eta)^{2}}$ with $\phi$ being the azimuthal angle and $\eta$
the pseudorapidity. The muon requirement is more efficient for
signal than background due to the presence of a $b\rightarrow\mu$ or
$b\rightarrow c\rightarrow\mu$ decay from at least one of the four
$b$ quarks, and is also required for an accurate measurement of the
trigger efficiency. Primary vertices (PVs) are reconstructed by
clustering tracks and correspond to \ppbar\ interaction locations.
To ensure good SV reconstruction, we further require fewer than four
PVs be reconstructed and that the selected PV with the largest
$\sum_{i}\log p_T^{i}$, summed over all vertex tracks $i$, be
located within $|z|<35$~cm and $r<1$~cm, where $x$ and $y$ are the
horizontal and vertical components of the distance $r$ with respect
to the beam axis, and $z$ is the distance along the beam axis from
the center of the detector. An initial selection requires that each
event has at least one SV with 2D decay length from the PV in the
plane transverse to the beam ($L^{xy}_d$) larger than 1~cm and decay
length significance (decay length divided by its uncertainty)
greater than five. The momentum of the SV, reconstructed from the
vectorial sum of the momenta of its associated tracks, must point
away from the PV to reduce combinatoric background. SVs are
reconstructed using a track selection so as to efficiently combine
the $b$ and $\bar{b}$ decay products of each v-hadron into a single
SV. Approximately 50 million data events satisfy these requirements,
dominated by dijet and heavy-flavor production.

To maximize the discovery potential of this analysis we use an OR of
all triggers. The most frequently fired triggers that make up the
dataset passing the initial selection involve a muon and jet at the
first trigger level and refinements of these objects at higher
levels. The overall trigger OR efficiency is estimated by first
measuring the efficiency for a single trigger per data collection
period using known muon and jet trigger efficiencies. Then the
number of events fired by that single trigger is compared to the
total number of data events passing the OR of all triggers, as a
function of sensitive variables, such as muon \pt, jet \pt, jet
angles, etc. No significant dependence is found, except on jet \pt,
thus the overall trigger OR efficiency is modeled as a function of
jet \pt.

Further selections are optimized by maximizing the expected signal
significance ($S/\sqrt{S+B}$), where $B$ and $S$ are the number of
MC background and signal events, respectively. The heavy-flavor
background, mainly $b$ hadrons with $c\tau$$\approx$0.3~cm, produces
a very large number of SVs, but their number decreases exponentially
as the radial distance of the SV from the PV increases. SVs are
required to have $L^{xy}_d$\gt1.6~cm. We expect signal events to
preferentially produce SVs with a large number of attached tracks,
therefore we require SVs with track multiplicity of at least four.
Interactions of primary collision particles ($\pi$, protons, etc.)
with detector material, such as silicon sensors, cables, etc., are
the major source of background. In order to quantify the material
regions, we construct a map of SV density in data, using SV with
track multiplicity of three, in the $xy$ (see
Fig.~\ref{fig:xy-material-map}) and $rz$ projections. SVs that occur
in regions of high SV density are then removed. After this
``pre-selection'' is performed, the multijet background MC sample is
normalized to the data (see Table~\ref{tab:pre-sel-summary}).
Finally, at least two SVs are required in each event, and they are
required to have $\Delta {\cal R}(SV1,SV2)$\gt 0.5, to prevent cases
where a single true vertex is mis-reconstructed as two nearby
separate vertices. No events in data have more than two SVs.

\begin{table}
\begin{centering}
\caption{\label{tab:pre-sel-summary}Summary of event selections,
showing the remaining background, data, and signal events (for
$M_H$=120~GeV and $L_d$=5~cm) after each selection. Background is
normalized to data after pre-selection.}
\begin{tabular}{ccccc}
\hline\hline
                           & $N_{\text{bkgd}}$ & $N_{\text{data}}$ & $m_{HV}$= & $m_{HV}$= \\
                           &                  &                  & 15~GeV    & 40~GeV    \\ \hline
 Produced                  & - &  -   & 2712 & 2712 \\
 Initial selection         & - & 4.9$\times10^7$  & 235 & 173 \\
 Trigger                   & - & 4.9$\times10^7$  & 174 & 77 \\
 SV $L^{xy}_d$\gt1.6~cm    & - & 3.2$\times10^7$  & 153 & 66 \\
 SV mult. $\geq$4          & - & 1.8$\times10^5$ & 72 & 25 \\
 SV density                & 6.0$\times10^4$ & 6.0$\times10^4$& 60 & 15 \\
\hline
 Num. SV $\geq$2           & 37.5 & 26& 5.1& 0.6 \\
\hline\hline
\end{tabular}
\par\end{centering}
\end{table}

\begin{figure}
\begin{centering}
\includegraphics[width=3.25in,height=2.5in]{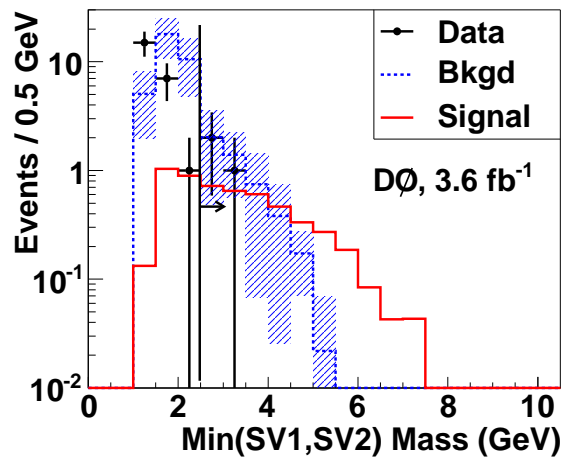}
\end{centering}
\caption{\label{fig:result-minmass-data-qcd-signal}The minimum mass
of the two SVs, for data, background MC, and signal MC with
$M_H$=120~GeV, $m_{HV}$=15~GeV, and $L_d$=5~cm. The hatched region
shows the uncertainty on the background MC.}
\end{figure}

\begin{figure}
\begin{centering}
\includegraphics[width=3.25in,height=2.5in]{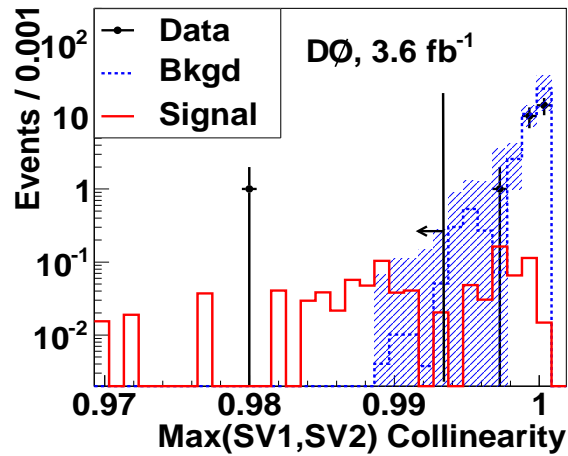}
\end{centering}
\caption{\label{fig:result-maxcoll-data-qcd-signal}The maximum
collinearity of the two SVs, for data, background MC, and signal MC
with $M_H$=120~GeV, $m_{HV}$=40~GeV, and $L_d$=5~cm. The hatched
region shows the uncertainty on the background MC.}
\end{figure}

Two more variables are used to select the signal: SV invariant mass
and SV collinearity. The invariant mass is reconstructed from the
four-momenta of the outgoing tracks attached to a SV, assuming the
pion mass for all particles. Collinearity is defined as the cosine
of the angle between the vector sum of the momenta of the attached
tracks and the direction to the SV from the PV. Depending on the
signal point, one of these two variables is used to perform the
final separation of signal and background. The quality of the
background model is of primary importance so we develop a method of
tuning the multijet background simulation of the SV invariant mass
and SV collinearity distributions to data. The events after
pre-selection are divided into two distinct sets: the first contains
events with only one SV (1SV), whereas the second contains events
with at least two SVs (2SV).
Since the signal content of the 1SV set is expected to be \lt0.1\%
for any of the signal MC points studied, we use the 1SV set to
compare the data to the multijet background MC and perform
corrections to the MC.
Gaussian smearing functions are applied to fractions of the
background MC events for the SV invariant mass and SV collinearity
to model the tails of the distributions better. The SV invariant
mass is smeared using a width of 12 GeV in about 1\% of events, and
the SV collinearity is smeared with a width of 0.15 in about 1.5\%
of events. The same smearing is then also applied to all MC signal
samples. For $m_{HV}<20$~GeV, a requirement on the minimum SV mass
in an event \gt 2.5~\gev\ is most effective
(Fig.~\ref{fig:result-minmass-data-qcd-signal}). For heavier
v-hadrons, we take advantage of the SV's decay products being more
widely spread in angle, which is better measured than the invariant
mass. Requiring the maximum SV collinearity in an event to be \lt
0.9937 maximizes the expected significance
(Fig.~\ref{fig:result-maxcoll-data-qcd-signal}).

The uncertainty on the signal acceptance is dominated by the
modeling of trigger efficiency and is (13--17)\%. The uncertainty on
the background due to the difference in track reconstruction
efficiency between MC and data is estimated by using two methods of
normalization and found to be 28\%. We estimate the effect of
smearing the MC samples by performing the entire analysis without
smearing.
For the requirements applied to the SV mass or collinearity,
smearing results in a difference of up to 18\% on the multijet
background yield and a negligible difference on the signal
acceptance.
Smearing also has no effect on the optimized requirement
values. To estimate the uncertainty from requiring a small SV
density, we compare the difference in the number of remaining events
between multijet background and data before and after making the
density requirement, and find agreement within (8--15)\%. The
uncertainty on the integrated luminosity is 6.1\% \cite{lumi}.

\begin{table*}
\begin{centering}
\caption{\label{tab:Summary-of-results} Results for each simulated
signal: the numbers of background, signal, and data events after all
selections, overall signal efficiency, SM Higgs production rate, and
observed and expected 95\% C.L.~upper limits on the signal cross
section.}
\begin{tabular}{ccccccccc}
\hline\hline
 $M_H$&$m_{HV}$&$L_d$& $N_{\text{bkgd}} \pm {\text{stat}} \pm {\text{sys}}$ & $N_{\text{sig}} \pm {\text{stat}} \pm {\text{sys}}$ & $N_{\text{data}}$ & Efficiency & SM Higgs (pb) & Limit obs. [exp.] (pb) \\
\hline
 90 GeV& 15 GeV& 5 cm    & $4.8\pm1.0\pm1.7$  & $3.3\pm0.3\pm0.5$   & 3 & 0.06\% & 2.0 & 3.2 [4.7] \\
 120 GeV& 15 GeV& 5 cm   & $4.8\pm1.0\pm1.7$  & $3.6\pm0.3\pm0.5$   & 3 & 0.13\% & 1.1 & 1.6 [2.4] \\
 120 GeV& 15 GeV& 2.5 cm & $4.8\pm1.0\pm1.7$  & $5.7\pm0.3\pm0.7$   & 3 & 0.21\% & 1.1 & 1.0 [1.5] \\
 120 GeV& 15 GeV& 10 cm  & $4.8\pm1.0\pm1.7$  & $1.5\pm0.2\pm0.3$  & 3 & 0.06\% & 1.1 & 3.9 [5.7] \\
 200 GeV& 15 GeV& 5 cm   & $4.8\pm1.0\pm1.7$  & $0.8\pm0.1\pm0.1$  & 3 & 0.16\% & 0.2 & 1.3 [1.8] \\
 90 GeV& 40 GeV& 5 cm    & $0.07\pm0.07\pm0.02$ & $0.15\pm0.07\pm0.03$ & 1 & 0.003\%& 2.0 & 67 [51]  \\
 120 GeV& 40 GeV& 5 cm   & $0.07\pm0.07\pm0.02$ & $0.38\pm0.07\pm0.06$ & 1 & 0.01\% & 1.1 & 16  [12] \\
 200 GeV& 40 GeV& 5 cm   & $0.07\pm0.07\pm0.02$ & $0.16\pm0.03\pm0.02$ & 1 & 0.03\% & 0.2 & 6.5 [5.1] \\
\hline\hline
\end{tabular}
\end{centering}
\end{table*}

\begin{figure}\centering
\includegraphics[width=3.2in]{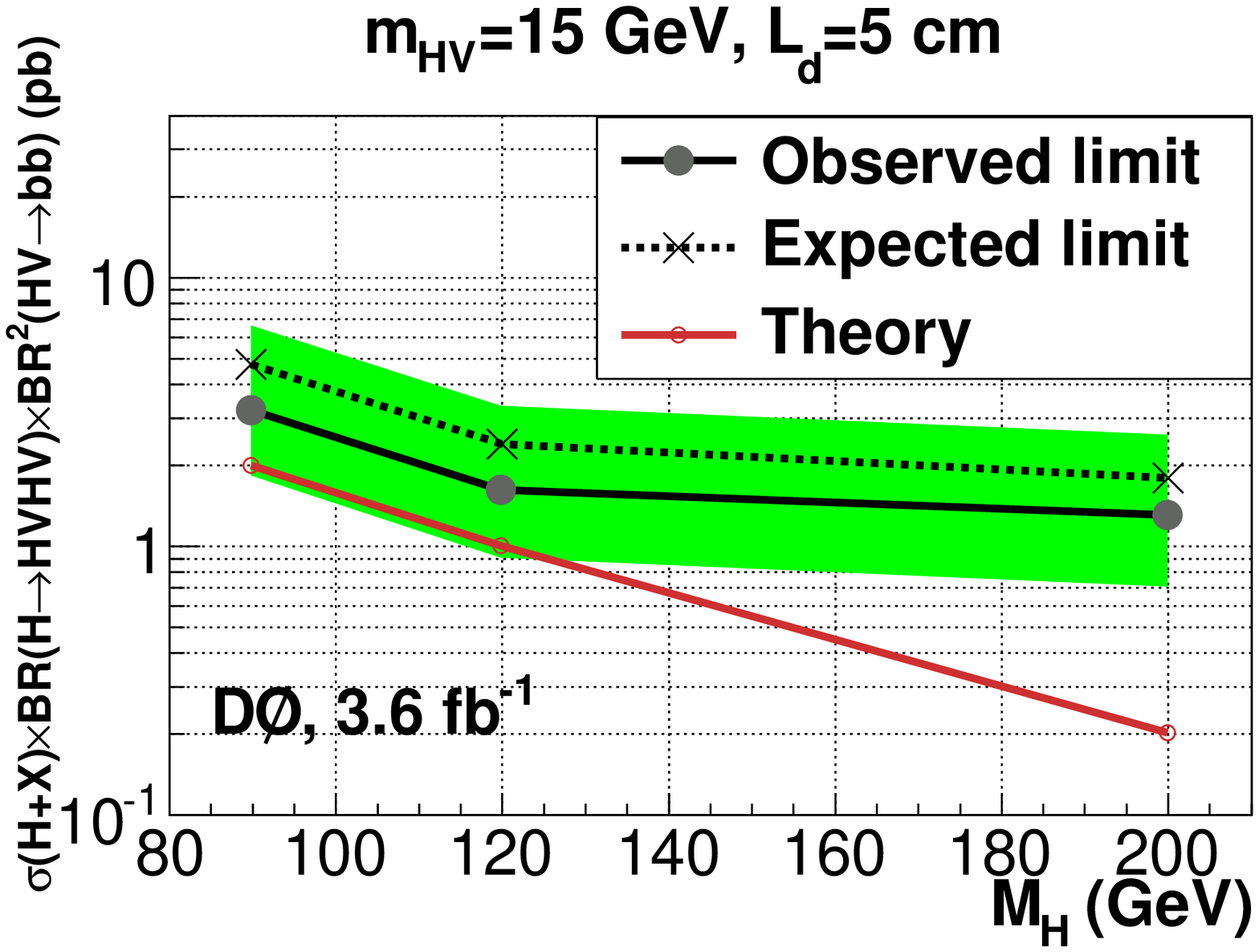}
\includegraphics[width=3.2in]{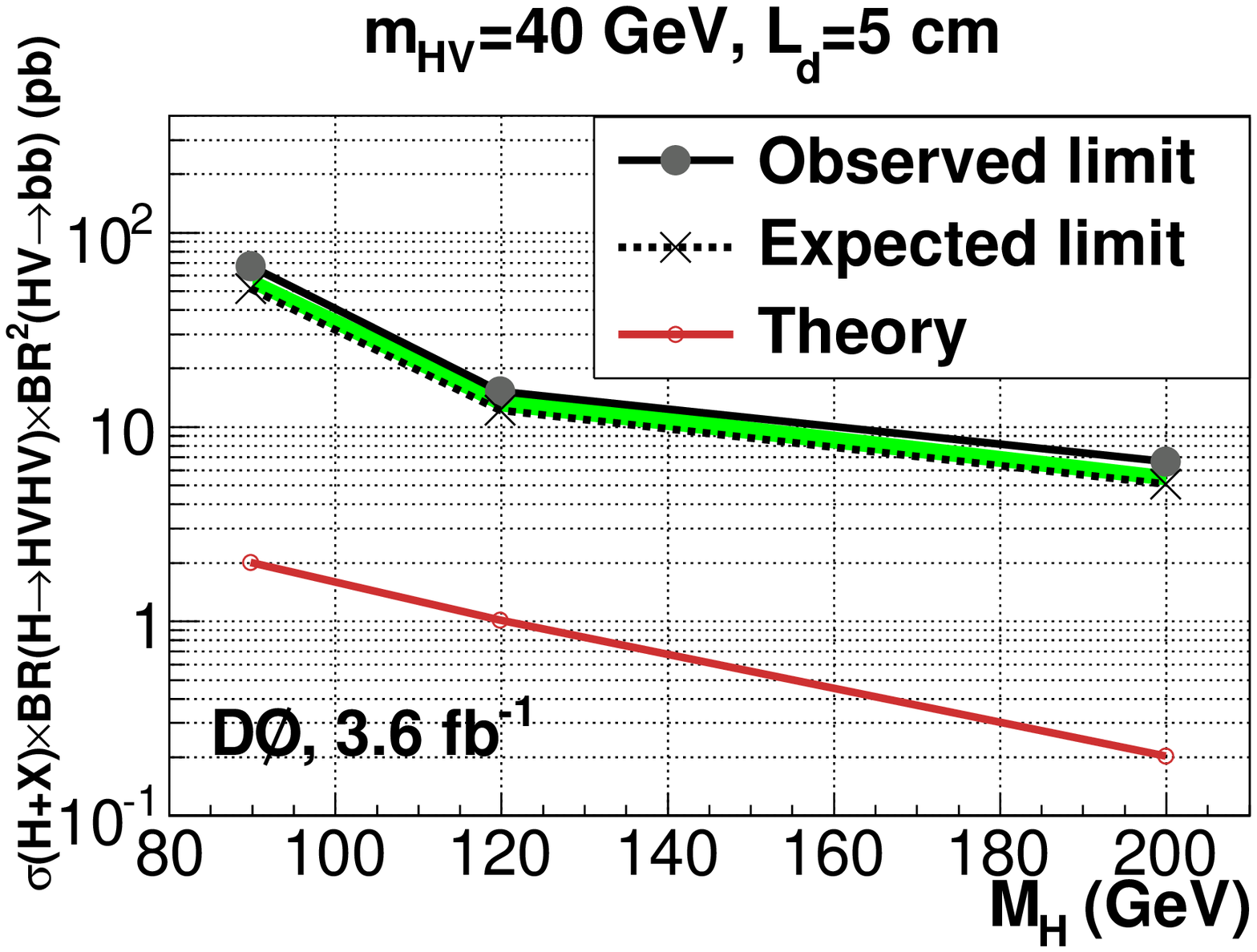}
\includegraphics[width=3.2in]{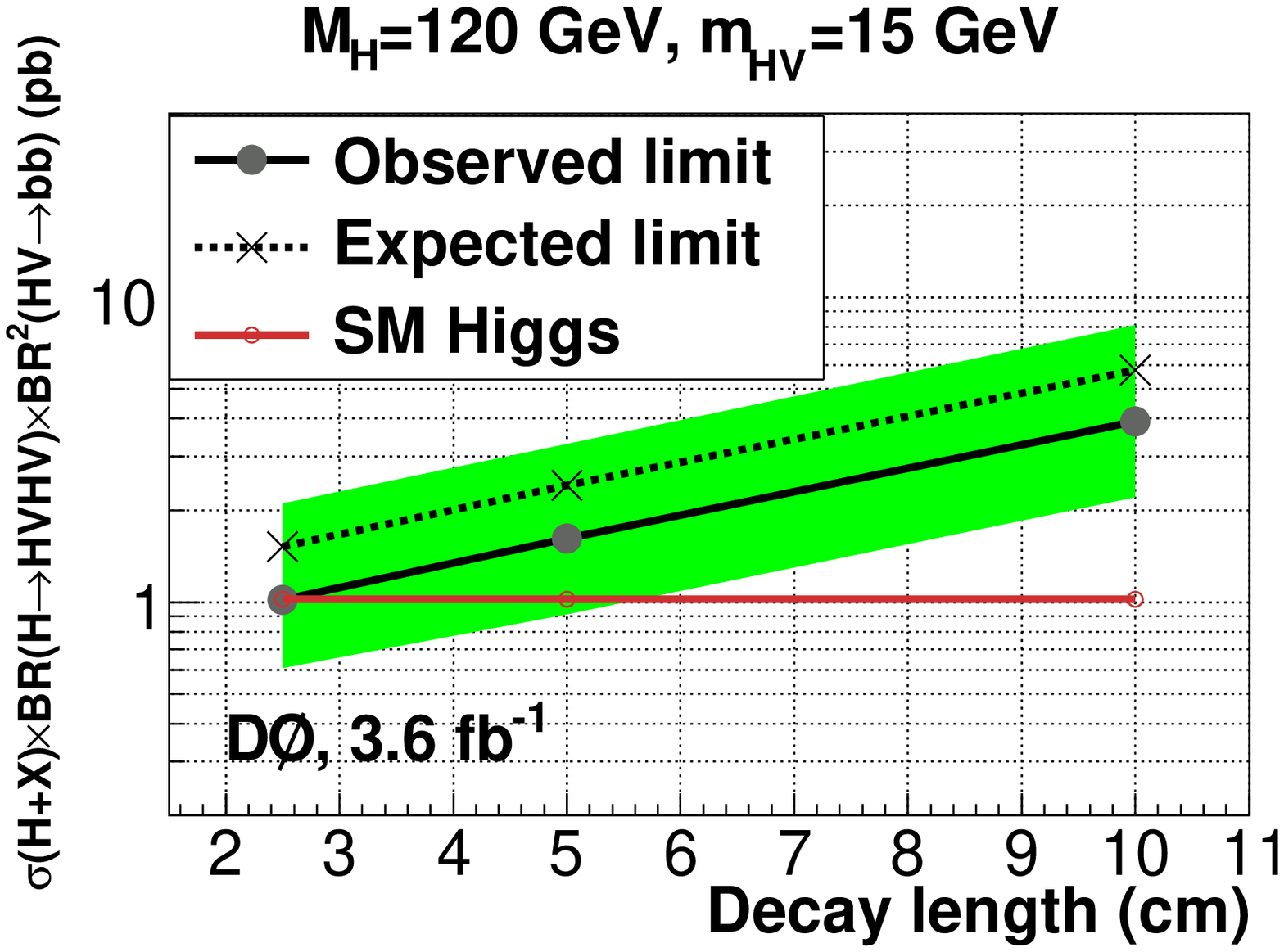}
\caption{The expected and observed 95\% C.L.~limits on
$\sigma$($H$+$X$)$\times$BR($H$\rarrow$HV
HV$)$\times$BR$^2$($HV$\rarrow\bbbar) for each $M_H$ studied,
$m_{HV}$ = 15, 40~\gev, and various values of v-hadron $L_d$. The
green band shows the $\pm$1 standard deviation on the expected
limit. The reference Higgs boson cross section from the
SM~\cite{Hahn:2006my} is shown by the solid red line, which assumes
100\% for BR($H$\rarrow$HV HV$) and BR($HV$\rarrow\bbbar). (color
online)} \label{fig:expobs}
\end{figure}

The final results after all selections are summarized in
Table~\ref{tab:Summary-of-results}. No significant excess is
observed, so 95\% C.L. limits are set on
$\sigma$($H$+$X$)$\times$BR($H$\rarrow$HV
HV$)$\times$BR$^2$($HV$\rarrow\bbbar) using a modified frequentist
method~\cite{topstat}, which includes all systematic uncertainties
on signal acceptance, background, and luminosity. Depending on the
signal parameters, Higgs boson production about 1--10 times the SM
cross section is excluded, if the Higgs boson always decays to a
pair of long-lived v-hadrons decaying only to \bbbar\ (see
Fig.~\ref{fig:expobs}). These results also provide the first
constraints on pair-produced NLLPs decaying to $b$ jets in the
radial range of 1.6--20~cm at a hadron collider.

%
We thank the staffs at Fermilab and collaborating institutions, 
and acknowledge support from the 
DOE and NSF (USA);
CEA and CNRS/IN2P3 (France);
FASI, Rosatom and RFBR (Russia);
CNPq, FAPERJ, FAPESP and FUNDUNESP (Brazil);
DAE and DST (India);
Colciencias (Colombia);
CONACyT (Mexico);
KRF and KOSEF (Korea);
CONICET and UBACyT (Argentina);
FOM (The Netherlands);
STFC and the Royal Society (United Kingdom);
MSMT and GACR (Czech Republic);
CRC Program, CFI, NSERC and WestGrid Project (Canada);
BMBF and DFG (Germany);
SFI (Ireland);
The Swedish Research Council (Sweden);
CAS and CNSF (China);
and the
Alexander von Humboldt Foundation (Germany).
%


\begin{thebibliography}{99}
%
\bibitem[a]{alton}
Visitor from Augustana College, Sioux Falls, SD, USA.
\bibitem[b]{askew,atramentov,gershtein}
Visitor from Rutgers University, Piscataway, NJ, USA.
\bibitem[c]{burdin}
Visitor from The University of Liverpool, Liverpool, UK.
\bibitem[d]{luna-garcia}
Visitor from Centro de Investigacion en Computacion - IPN,
  Mexico City, Mexico.
\bibitem[e]{podesta-lerma}
Visitor from ECFM, Universidad Autonoma de Sinaloa, Culiac\'an, Mexico.
\bibitem[f]{voutilainen}
Visitor from Helsinki Institute of Physics, Helsinki, Finland.
\bibitem[g]{weber}
Visitor from Universit{\"a}t Bern, Bern, Switzerland.
\bibitem[h]{wenger}
Visitor from Universit{\"a}t Z{\"u}rich, Z{\"u}rich, Switzerland.
\bibitem[\ddag]{deceased}
Deceased.

%
\vskip 0.25cm

\bibitem{strass2006_1}
M.~J.~Strassler and K.~M.~Zurek,
  Phys.\ Lett.\  B {\bf 651}, 374 (2007).

\bibitem{strass2006_2}
  M.~J.~Strassler and K.~M.~Zurek,
  Phys.\ Lett.\  B {\bf 661}, 263 (2008).

\bibitem{lep2003}
 R.~Barate {\it et al.},
 Phys.\ Lett.\ B {\bf 565}, 61 (2003).

\bibitem{spencer_higgs_decays}
S.~Chang, R.~Dermisek, J.~F.~Gunion, and N.~Weiner,
  Ann.\ Rev.\ Nucl.\ Part.\ Sci.\  {\bf 58}, 75 (2008).

\bibitem{d0det}
 D0 Collaboration, V. Abazov {\it et al.}, Nucl. Instrum. Methods Phys.\
 Res.\ A. {\bf 565}, 463 (2006).

\bibitem{pythia}
 T.~Sj\"ostrand {\it et al.}, Comput.\ Phys.\ Commun.\ {\bf 135}, 238
(2001).

\bibitem{geant}
 R. Brun and F. Carminati, CERN Program Library Long Writeup W5013,
1993 (unpublished).

\bibitem{RunIIcone}
G.~C.~Blazey {\it et al.}, arXiv:hep-ex/0005012 (2000).

\bibitem{lumi}
 T. Andeen {\it et al.}, FERMILAB-TM-2365 (2007).

\bibitem{Hahn:2006my}
  D.~de Florian and M.~Grazzini,
  Phys.\ Lett.\  B {\bf 674}, 291 (2009).

\bibitem{topstat}
T.~Junk, Nucl. Instrum.\ Methods A. {\bf 434}, 435 (1999).

\end{thebibliography}
\end{document}